\documentclass[12pt,preprint]{aastex}

\newcommand\msun{M_{\odot}}

\newcommand\be{\begin{equation}}
\newcommand\en{\end{equation}}

\newcommand\etal{{\rm et al}.\ }

\def\micron{$\mu$m}
\def\microns{$\mu$m }

\begin{document}

\shortauthors{Muzerolle et al.}
\shorttitle{Spitzer census of transition disks}

\title{A {\it Spitzer} Census of Transitional Protoplanetary Disks with AU-Scale Inner Holes}

\author{
James Muzerolle\altaffilmark{1,2},
Lori E. Allen\altaffilmark{3},
S. Thomas Megeath\altaffilmark{4},
Jes\'us Hern\'andez\altaffilmark{5,6}
Robert A. Gutermuth\altaffilmark{7}
}

\altaffiltext{1}{Space Telescope Science Institute, 3700 San Martin Dr., Baltimore, MD 21218}
\altaffiltext{2}{Steward Observatory, 933 N. Cherry Ave., The University of Arizona, Tucson, AZ 85721}
\altaffiltext{3}{National Optical Astronomy Observatory, 950 N. Cherry Ave., Tucson, AZ 85719}
\altaffiltext{4}{Ritter Observatory, Department of Physics and Astronomy, University of Toledo, 2801 W. Bancroft Ave., Toledo, OH 43606}
\altaffiltext{5}{Department of Astronomy, University of Michigan, 830 Dennison Building, 500 Church Street, Ann Arbor, MI 48109}
\altaffiltext{6}{Centro de Investigaciones de Astronom\'ia, Apdo. Postal 264, M\'erida 5101-A, Venezuela}
\altaffiltext{7}{Five Colleges Astronomy Dept., Smith College, Northampton, MA  01063}

\begin{abstract}
Protoplanetary disks with AU-scale inner clearings, often referred to as
transitional disks, provide a unique sample for understanding disk dissipation
mechanisms and possible connections to planet formation.  Observations of
young stellar clusters with the {\it Spitzer} Space Telescope have amassed
mid-infrared spectral energy distributions for thousands of star-disk systems
from which transition disks can be identified.  From a sample of 8 relatively
nearby young regions ($d \lesssim 400$ pc), we have identified about 20
such objects, which we term ``classical" transition disks, spanning a wide
range of stellar age and mass.
We employed strict infrared continuum criteria to limit ambiguity:
an 8 to 24 \microns spectral slope limit ($\alpha > 0$)
to select for robust optically thick outer disks, and 3.6 to 5.8 \microns
spectral slope and 5.8 \microns continuum excess limits to select for
optically thin or zero continuum excess from the inner few AU of the disks.
We also identified two additional categories representing more ambiguous cases: 
"warm excess" objects with transition-like spectral energy distributions
but moderate excess at 5.8 \micron, and "weak excess" objects with
smaller 24 \microns excess
that may be optically thin or exhibit advanced dust grain growth and settling.
From existing H$\alpha$ emission measurements, we find evidence for
different accretion activity among the three categories, with a majority of
the classical and warm excess transition objects still accreting gas
through their inner holes and onto the central stars, while a smaller
fraction of the weak transition objects are accreting at detectable rates.
We find a possible age dependence to the frequency of classical transition
objects, with fractions relative to the total population of disks in a given
region of a few percent at 1-2 Myr rising to 10-20\% at 3-10 Myr.  The trend
is even stronger if the weak and warm excess objects are included.
This relationship may be due to a dependence of the outer disk clearing
timescale with stellar age, suggesting a variety of clearing mechanisms
working at different times, or it may reflect that a smaller fraction of
all disks actually undergo an inner clearing phase at younger ages.
Classical transition disks appear to be less common, and weak transition
disks more common, around lower-mass stars ($M \lesssim 0.3 \msun$),
which we suggest may be a further indicator of the stellar mass-dependent
disk evolution that has been seen in previous studies.
The difference in number statistics and accretion activity between
the two classes further suggests that they are not connected but rather
represent distinct evolutionary outcomes for disks.

\end{abstract}


\section{Introduction}

The lifetimes of circumstellar disks around young stars determine
the relevant timescales available for planet formation to occur.
Observational estimates are essential for constraining theories
such as the core accretion (Pollack et al. 1996) and gravitational
instability (Boss 1997) models.  A considerable amount of work
has gone into estimating
disk lifetimes primarily by examining infrared excess emission
as a function of age.  Previous studies focused on near-infrared emission
probing the innermost regions of disks have indicated typical disk
lifetimes of roughly 3 Myr, with a wide dispersion from 1-10 Myr
(Strom et al. 1989; Haisch et al. 2001; Hillenbrand 2005).

Constraints on the lifetime of disk material in the planet formation
zone of 0.1-10 AU have been more difficult to achieve because of
sensitivity and resolution limitations at the longer wavelengths
which probe cooler dust.  Early results from ground-based and IRAS
observations of the nearby Taurus star forming region by
Skrutskie et al. (1990) first identified a small population of
disks ``in transition", lacking excess emission at
$\lambda \lesssim 10$ \microns but exhibiting considerable emission
at longer wavelengths and thus implying a lack of dust (at least in
small grains) in the inner few AU of the disks.  Given the age
of Taurus, Skrutskie et al. estimated a typical transition timescale
of $\sim 0.3$ Myr.

A handful of such transition disks with inner ``holes"
have been studied in some detail in recent years
\footnote{The ``transition disk" terminology is somewhat subjective
and has a wide range of definitions in the literature.  For the purposes
of this paper, we define a ``classical" transition disk specifically as an
optically thick protoplanetary disk with an $\sim$ AU-scale
inner region that is optically thin or completely evacuated of small
dust grains.  We alternately refer to these as disks with inner holes.}.
Calvet et al. (2002) modeled the spectral energy distribution (SED)
of the 10 Myr-old classical T Tauri star (CTTS) TW Hya with an inner disk
hole of size $\sim 4$ AU,
and hypothesized that a Jupiter-mass planet may have formed there.
Similar conclusions have been drawn for the Taurus objects
GM Aur (Rice et al. 2003; Bergin et al. 2004; Calvet et al. 2006),
DM Tau (Bergin et al. 2004; Calvet et al. 2006) and CoKu Tau/4
(Forrest et al. 2004; Quillen et al. 2004; D'Alessio et al. 2005),
though the latter has subsequently been found to harbor a stellar
companion (Ireland \& Kraus 2008).  Follow-up observations at high
spatial resolution have directly confirmed the presence of holes or gaps
in a few disks (Hughes et al. 2007; Ratzka et al. 2007; Brown et al. 2008;
Hughes et al. 2009).
However, the overall frequency of transition objects as a function of
stellar age, mass, and environment has remained largely unconstrained.

With its superior mid-infrared sensitivity and spatial resolution,
the {\it Spitzer} Space Telescope has begun to address questions of
disk evolution and lifetimes and the connection to planet formation
in unprecedented detail.  {\it Spitzer} observations taken with the IRAC and
MIPS instruments are providing SEDs from
3.6 to 24 \microns for large numbers of pre-main sequence stars in
star forming regions and young clusters, allowing statistical studies of
circumstellar disk evolution through the properties of dust emission
as a function of stellar age, mass, and environment (Lada et al. 2006;
Sicilia-Aguilar et al. 2006; Hern\'andez et al. 2007, Balog et al. 2007;
Kennedy \& Kenyon 2009, among many others).  Here we report on the statistics
of transition disks as identified from Spitzer SEDs for stars in nearby
young regions, providing the first robust
statistics of disks with inner holes around stars
in young stellar clusters and associations with ages 1-10 Myr.

\section{Observations and Sample}

We have culled disk candidates from 7 regions observed as part of
existing {\it Spitzer} GTO and GO surveys of young clusters and star forming
regions.  The sample is listed in Table~\ref{clusters}.
We included only regions within about 400 pc so as to optimize
completeness limits, particularly at longer wavelengths, and restricted
to mean stellar ages of 10 Myr or less.  The sample includes four clusters
with active star formation, and three older stellar associations
where star formation has ceased (and the natal molecular material has
been swept away).  Some transition objects in some of these regions
have already been reported in the literature; however, the nomenclature
and methodology varies widely.  Our aim here is to narrowly define
transition disks and provide complete statistics by using a consistent
selection method for identifying them.

Each region was mapped with both the IRAC and MIPS instruments,
with total coverage ranging from 30'x30' to 60'x60'.  Data from each
instrument are separated in time by a few weeks to a few months;
however, all channels within each instrument were obtained simultaneously.
We defer the details of data processing, mosaicking, and photometry
to separate papers on each of the regions, referenced in Table~\ref{clusters}.
We do not consider data from the 70 or 160 \microns channels of MIPS here
because of severe incompleteness and saturation effects.
The {\it Spitzer} sources were matched with 2MASS data, and the 2MASS
coordinates were adopted in each case (all the disk candidates
discussed here have a matching 2MASS source detected at $J$, $H$, and $K$).

\section{Results}

\subsection{SED classification}

\begin{figure}
\plotone{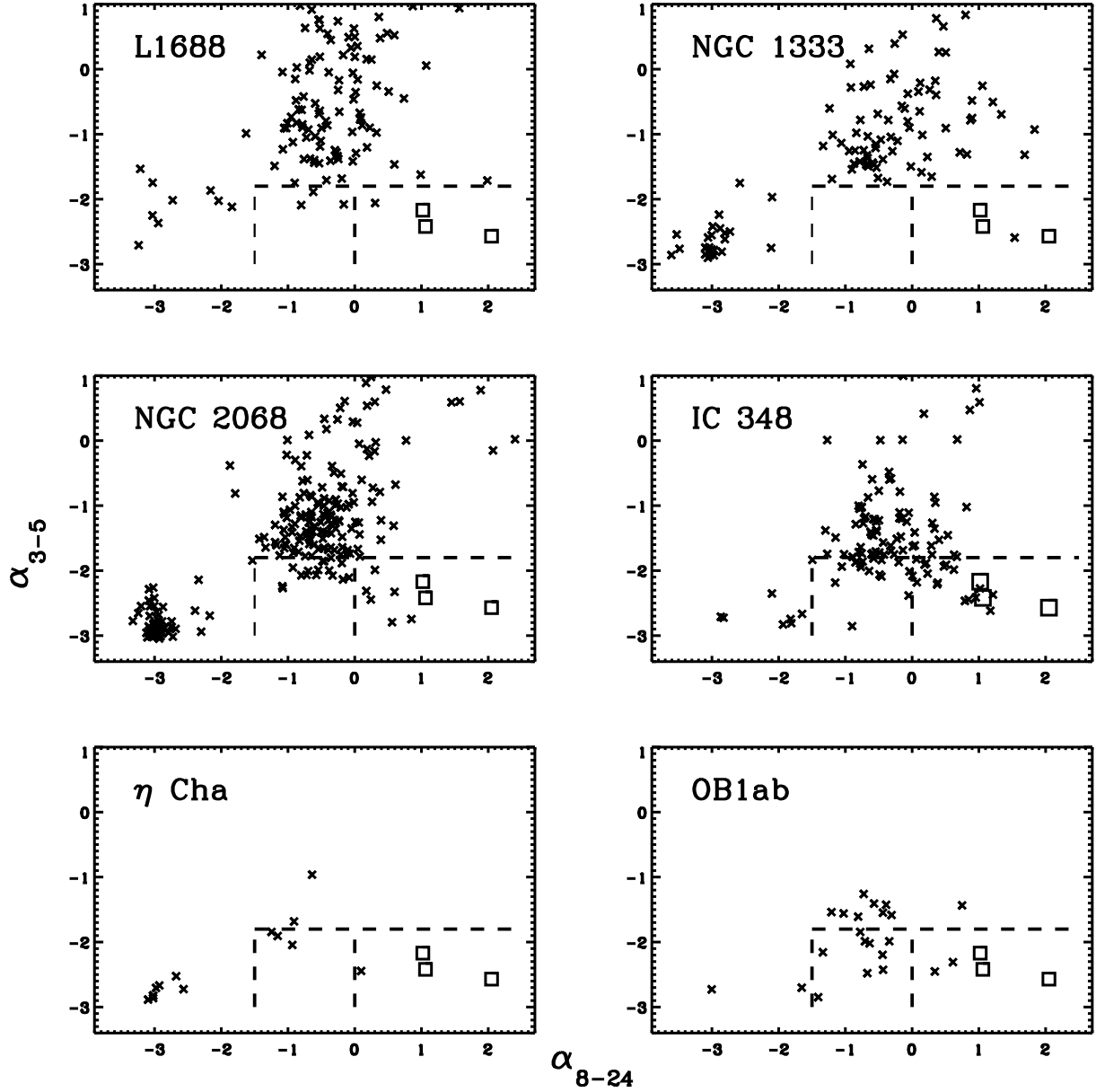}
\caption{IRAC-MIPS spectral slope comparison.  {\it Crosses:} all objects
detected at 3.6, 4.5, 5.8, 8, and 24 $\mu$m in L 1688, NGC 1333, and NGC 2068
(some very red objects are off the top edge of the plots), and confirmed
members only for IC 348, the $\eta$ Cha association and the Orion OB1a and b
associations.
{\it Squares:} the known transition
objects GM Aur, TW Hya, and CoKu Tau/4 (from left to right).
Dashed lines mark the adopted region for selecting
transition disk candidates (see text).
\label{specslope}}
\end{figure}

We have selected circumstellar disk candidates from each region
based on spectral slope measurements across the IRAC and MIPS 24
\microns channels.  The spectral slope provides a useful means of
distinguishing between stars with and without infrared excess emission
from circumstellar dust, and can also be used to distinguish rough
evolutionary characteristics (e.g. Lada et al. 2006; Hern\'andez et al. 2007;
Flaherty \& Muzerolle 2008).  We focused on two different wavelength intervals,
3 to 5.8 \microns and 8 to 24 \micron, in order to sample dust at two
different temperature regimes (and hence locations in the disk).
Figure~\ref{specslope} shows these slope values for the 7 regions
in our sample (data from the Orion OB1a and b associations have been
combined in one panel).  Sources clustered at $\alpha_{8-24} \sim -3$
represent bare photospheres with little or no excess dust emission.
The locus of points in the range $0 \le \alpha_{8-24} \le -1$,
$-1.8 \le \alpha_{3-5} \le 0$ are objects with optically thick
circumstellar disks.  Many of these have been confirmed with follow-up
optical spectroscopy as being T Tauri stars.

We lack complete membership confirmation
for three of the clusters (NGC 1333, L 1688, NGC 2068/2071).
Other non-PMS objects can exhibit similar spectral slopes from dust excess,
mainly high-redshift galaxies and AGN and background AGB stars.
Since the vast majority of background objects are faint, we attempted
to limit the contamination for those three clusters by applying
a magnitude cut of $[3.6]<14.5$.
We applied a further cut using $[4.5]-[8]$ and $[5.8]-[8]$
color criteria from Winston et al. (2007) and Gutermuth et al. (2008)
to remove star forming galaxies and AGN with strong PAH emission.
Based on extragalactic source counts and observations of galactic fields
which lack young stellar objects,
we estimate the final level of contamination in the disk locus
to be very low, of order 1\% or less (see also Gutermuth et al. 2008).
In any case, contamination is not an issue for the other four regions,
which have extensive published membership information,
and we consider confirmed members only.

It is apparent in Figure~\ref{specslope} that there are a handful
of sources with flat or increasing slopes from 8-24 \microns but sharply
declining (near-photospheric) slopes at 3-5 \micron.  Such behavior is
indicative of optically thick dust emission at the longer wavelengths but
very small or no dust excess at the shorter wavelengths.
For typical T Tauri-type disks, emission at 4.5 \microns probes dust
at distances of $\sim 1$ AU from the central star.
Thus, the lack of excess emission
at the shorter wavelengths points to an optically thin or evacuated
(at least of small grains) ``hole" in the disk at those radii,
while optically thick material exists further out at $\gtrsim 1$ AU.
For comparison, the spectral slopes for three well-studied transition disks
(see Table~\ref{known})
are also shown in Figure~\ref{specslope}.  For selecting inner disk hole
candidates, we then adopted a spectral slope locus based on these
previously known objects, with thresholds $\alpha_{3-5} < -1.8$ and
$\alpha_{8-24} > 0$.  The former criterion roughly
corresponds to the maximum slope value for reddened background stars,
and is also similar to values applied previously to separate
``normal" accretion disk emission from more evolved or optically thin
disk emission (e.g. Lada et al. 2006; Hern\'andez et al. 2007; Flaherty \&
Muzerolle 2008).  Those studies calculated slopes across the full IRAC
wavelength range, 3.6 to 8 \micron.  We have excluded the 8 \microns band
in order to avoid artificially large slopes due to strong
10 \microns silicate emission, which partially overlaps the 8 \microns
bandpass, as well as potential contamination from PAH emission in the
surrounding molecular cloud material.  The longer wavelength criterion
was selected to weed out potentially optically thin outer disks;
$\alpha_{8-24} = 0$ corresponds to $L_{IR}/L_* \sim 0.05$
in the limit of no 8 \microns excess and single-temperature blackbody
dust emission.  This cutoff also avoids completeness limitations at 24 \micron.

\begin{figure}
\epsscale{0.7}
\plotone{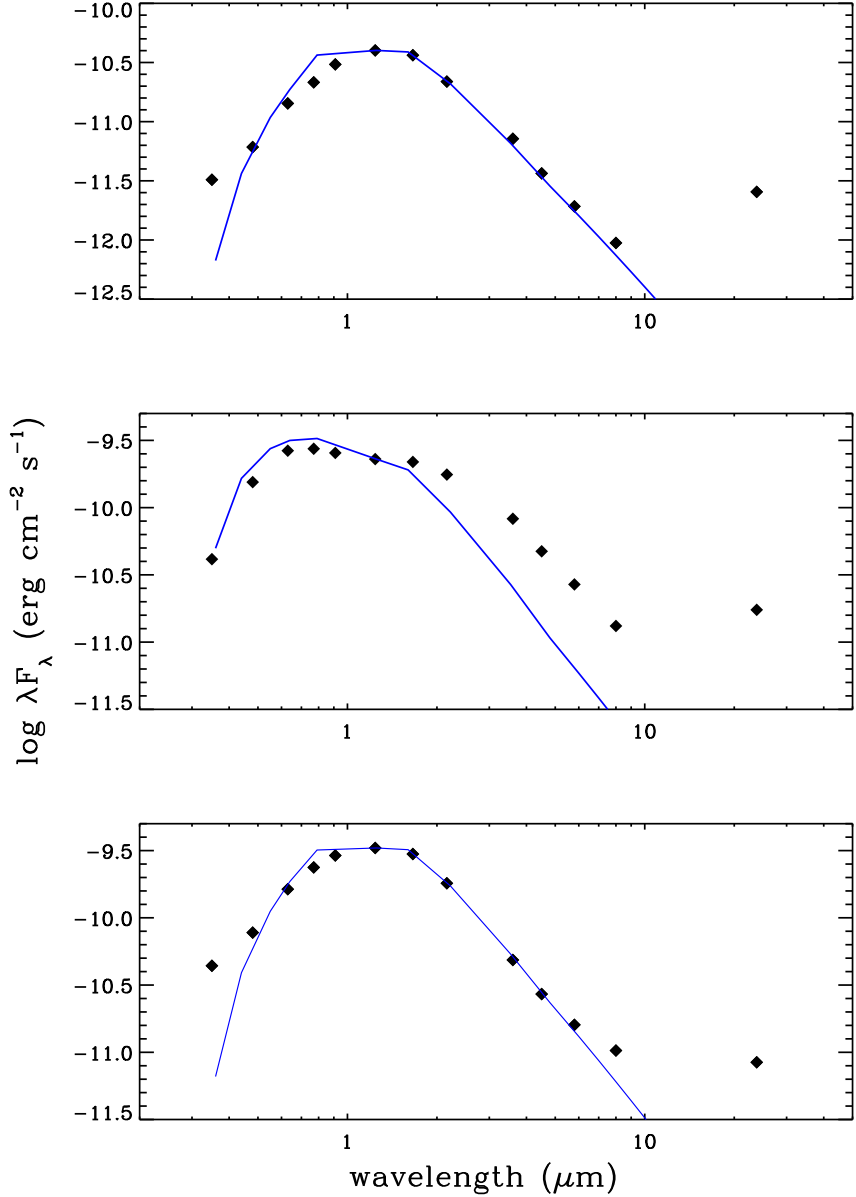}
\caption{SEDs of two inner disk hole candidates identified from
the spectral slope locus shown in Figure~\ref{specslope} (top two panels),
plus one object satisfying only the short wavelength criterion (bottom panel).
Diamonds are observed fluxes dereddened using the known $A_V$
of each star and the Mathis (1990) and Flaherty et al. (2007) reddening law.
Optical and infrared
photometry and stellar properties are reported in Flaherty \& Muzerolle (2008).
Solid blue lines are empirical photospheres constructed from
the main sequence star colors from Kenyon \& Hartmann (1995),
normalized to the observed $J$-band flux.
\label{seds}}
\end{figure}

SEDs for two transition disk candidates, plus a third object that satisfies
only the short wavelength criterion, are shown in Figure~\ref{seds}.
The object in the top panel exhibits nearly pure photospheric emission
out to 5.8 \micron, and a sharp upturn from 8 to 24 \micron, as expected.
On the other hand, the object in the second panel shows a similar SED shape
but has a clear excess
above the predicted photosphere at wavelengths as short as 2 \microns
despite having a steep 3-5 \microns spectral index.  In this case,
the excess emission at $\lambda < 8$ \microns is consistent with
a single-temperature blackbody of $\sim 1400$ K (probably dominated by
emission from the disk ``wall" at the dust sublimation radius;
e.g. Muzerolle et al. 2003; Espaillat et al. 2008a), which the spectral
index technique is not able
to distinguish from blackbody emission typical of a stellar photosphere.
The short-wavelength excess emission is significantly reduced from
``typical" T Tauri disks, akin to the ``evolved" disks reported by various
authors (e.g., Lada et al. 2006; Hern\'andez et al. 2007).
However, the interpretation of its origin is more complicated than
in the case of truly optically thin or zero excess emission as seen in
the top panel of Figure~\ref{seds}, particularly for the lowest mass stars
where hot dust generally produces less emission (Ercolano et al. 2009).
Finally, the object in the lower panel exhibits no dust emission at
$\lambda < 5.8$ \microns but weaker excess at 24 \microns with no obvious
upturn in the long-wavelength continuum.  Objects with this type of SED
have much lower fractional infrared luminosities, and cannot be reliably
distinguished from fully optically thin disks (such as debris disks)
with our data alone.

We have attempted to eliminate ambiguous cases from our statistics
by estimating the actual amount of excess at 5.8 \micron.
This was accomplished by comparing to a predicted photospheric SED
and calculating the ratio of observed to predicted flux at the wavelength
of interest.  In this case, we constructed empirical SEDs as a function
of spectral type from the main sequence dwarf colors given by Kenyon \&
Hartmann (1995).  Such a procedure is sensitive to both the spectral type
of the object and the amount of extinction it suffers.  We culled
spectral types and $A_V$ values for all transition candidates
from the literature (see Tables~\ref{holes}-~\ref{weak24}).
Each object SED was dereddened using these $A_V$ values
and reddening laws from Flaherty et al. (2007) (for $\lambda = 3.6-8$ \micron)
and Mathis (1990) (all other wavelengths).  The observed and photospheric
SEDs were then normalized at $J$ where excess emission from dust should
be negligible, particularly for these evolved disk candidates.  In some cases,
the literature $A_V$ values resulted in a poor match between observed
and photospheric SED in the near-infrared, with the observed fluxes
being systematically lower.  In these cases, we used a $\chi$-squared
procedure to determine the $A_V$ value that produced the best fit between
the observed $JHK$ fluxes and the photosphere template
(the change from published values was never more than $\sim 30\%$).
The predicted photospheric flux at 5.8 \microns was then calculated by
convolving the IRAC channel-3 bandpass function with a spline fit to
the scaled empirical SED.

\begin{figure}
\plotone{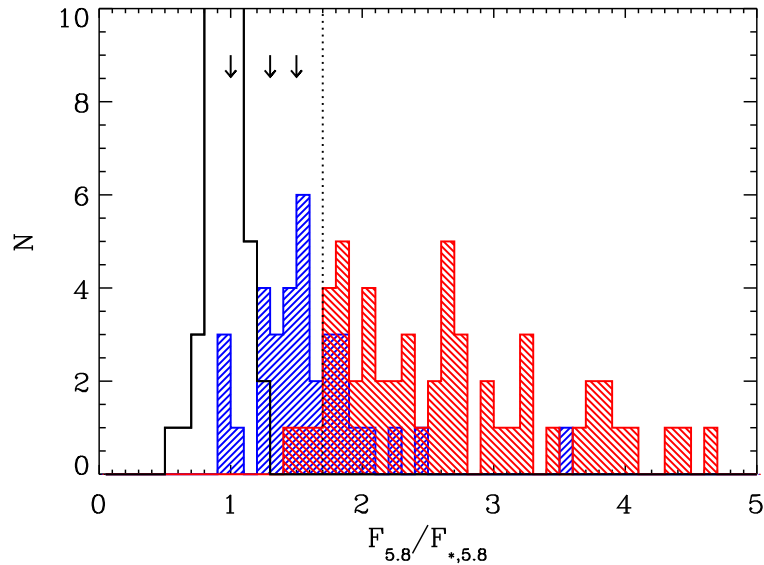}
\caption{The distribution of observed to photospheric flux ratios
at 5.8 \microns for IC 348 members.  The open histogram shows flux
ratios for all objects whose 3-5 \microns spectral slopes and lack
of 24 \microns detections indicate no substantial circumstellar material
(the peak is off the plot at $N=56$).
The hatched histograms represent flux ratios for sources with disks
and detections out to 24 \micron: positive-slope hatching represents
objects with $\alpha_{3-5} < -1.8$, negative-slope hatching represents
all other disks.  There are 8 objects with flux ratios from 5-15 not shown.
The arrows indicate the flux ratios of three known classical transition disks
(from left to right: CoKu Tau/4, TW Hya, GM Aur).
The dotted line indicates the cut-off value for our final
transition disk selection.
\label{fluxratio}}
\end{figure}

Figure~\ref{fluxratio} shows the distribution of observed to predicted
5.8 \microns flux ratios ($f_{5.8}$) for three subsets of the members of
the IC 348 cluster.  All objects with no evidence for infrared excess emission
at any wavelength based on the spectral slope are distributed around
$f_{5.8} \sim 1$, as expected.
Objects with 24 \microns excess but $\alpha_{3-5} < -1.8$ show a broader
distribution that peaks around $f_{5.8} \sim 1.5$.  This result indicates
that most of the transition disks do in fact have non-zero
excess emission at this wavelength, probably from optically thin dust
(see discussion).  By contrast, all but 3 of the objects with
24 \microns excess and $\alpha_{3-5} > -1.8$ show much larger flux ratios,
again as expected.
\footnote{However, we note that the equivalent value of this flux ratio
for the Taurus median SED (D'Alessio et al. 1999),
assuming a K7 photosphere, is about 5.4.  That almost all of the IC 348
``normal" disks are below this value further indicates the overall
advanced state of disk evolution for this older cluster, as first shown
by Lada et al. 2006.}
By comparing to the flux ratios of the three Taurus reference
transition disks, and adopting a reasonable boundary between
the candidate transition and normal disk flux ratio distributions
in Figure~\ref{fluxratio}, we added a final transition disk selection
constraint of $f_{5.8} < 1.7$.  The final list of classical transition
disks is shown in Table~\ref{holes}.  The remaining candidates that
did not pass the $f_{5.8}$ criterion, indicating larger excess emission
in the IRAC wavelength range, are listed in Table~\ref{marginal}
as ``warm excess" transition disks.
Finally, remaining objects that passed the $f_{5.8}$ criterion but exhibited
weaker 24 \microns excess emission ($\alpha_{8-24}<0$) are listed in
Table~\ref{weak24} as ``weak excess" transition disks.  These last two sets
of objects are discussed separately in section 3.3.

\subsection{Statistics}

We now estimate the frequency of disks with AU-scale inner holes
in each of the sample regions.
The stellar populations have not been completely characterized 
in all cases, so a frequency relative to the total number of stars
$N_*$ cannot be calculated reliably.  However, the total number of objects
in each region with excess emission out to 24 \microns characteristic of
optically thick circumstellar disks can be estimated
from the {\it Spitzer} spectral slopes alone.  Since contamination
from unrelated sources is very low, as estimated above, this number
should be a reliable indicator of the total number of protoplanetary disks
$N_{disk}$ in each region (at least for most of the stellar mass range;
see below).  The resulting classical transition disk frequency, expressed as
$N_{hole}/N_{disk}$, is given for each region in Table~\ref{clusters}.
The fraction of disks with inner holes ranges from about 1\%
in the youngest, most embedded clusters to 17\% in the oldest regions.

Our statistics are potentially biased since the sensitivity at 24 \microns
is not sufficient to detect photospheres for any but the earliest
spectral types at the distances of the clusters in the sample.
In addition, the bright background emission at 24 \microns endemic
to star formation regions further limits sensitivity, though
in a nonuniform manner.  The limiting case of a transition disk
with no excess at 8 \microns, a spectral slope $\alpha_{8-24} = 0$,
and a 24 \microns flux at the typical completeness limit $[24] \lesssim 9$
roughly corresponds to a 1 Myr-old M5 star at 320 pc.
The 24 \microns limit will be brighter in small regions where the background
is strongest and most highly structured, though the effect should
be relatively random within a given region since there is no direct
correlation between the background emission and the cluster spatial
distributions.  However, the overall background does vary among
the regions of our study.  The average brightness and peak-to-valley
variations are the most significant in IC 348 and L 1688, somewhat
more moderate in NGC 2068/2071, bright but uniform in NGC 1333,
and very low in the three older regions.

We can make a more detailed assessment of the completeness with IC 348,
which has good number statistics, complete membership, and an intermediate
distance for our sample.
For all the sources detected at all IRAC bands but not at 24 \micron,
we selected remaining disk candidates from sources with the appropriate
$\alpha_{3-5}$.  There are also many sources with possible excess at only
8 \micron.  We selected those with the criteria $[5.8]-[8] > 0.1$
and $[5.8]-[8] > 3 \sigma_{5.8-8}$.  The former cut corresponds to
the smallest color in our transition sample, while the latter cut
eliminates objects whose photometry has large uncertainties because of
strong background emission.  We also eliminated all objects with
spectral types later than M6.5 since none have 24 \microns
detections.  From these cuts we identified a total of 35 objects
with likely excess emission at 8 \microns or less.  Of these, 6 exhibit
$f_{5.8}$ and 24 \microns upper limits consistent with being transition
disks (though of course many of these could be in our weak excess
category).  All but two have spectral types M5 or later.
Combining with the 24 \microns-detected sample, we get a total of
134 objects with circumstellar disks, of which anywhere from 12 to 18
may exhibit mid-IR emission characteristic of optically thick outer disks
with AU-scale inner holes.  This yields a range in the transition disk
fraction of 9 - 13\%, within the 1$\sigma$ confidence interval of our
original estimate.  We conclude that the statistics we have derived from
the 24 \microns-detected samples are not significantly biased,
and are probably accurate within the uncertainties assuming binomial
counting statistics.

\subsection{Transition disk properties}

\subsubsection{Stellar properties and environment}

\begin{figure}
\plotone{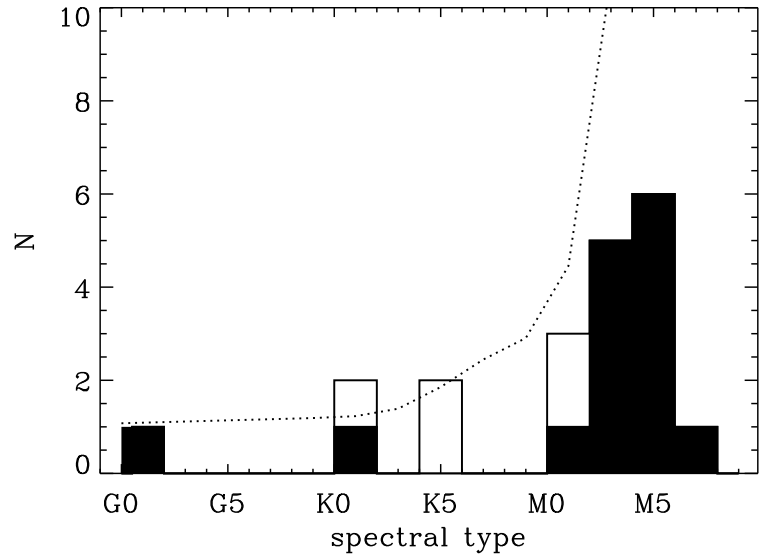}
\caption{Distribution of spectral types for the classical transition objects
identified in this paper.
The solid histogram shows the distribution for the transition disks
in the four oldest regions (IC 348, $\eta$ Cha, and Orion OB1ab).
The stellar mass function from Kroupa (2001), arbitrarily scaled to G0,
is shown with the dotted line.
\label{spt}}
\end{figure}

We have spectroscopic information for all 20 of our classical transition disks
(Table~\ref{holes}).  All of them are confirmed to be {\it bona fide}
pre-main sequence members of their respective clusters/associations
on the basis of spectral type, Li absorption, hydrogen line emission,
and/or various gravity-sensitive photospheric features (see for instance
Luhman et al. 2003 on IC 348 for details of youth and membership criteria).
The candidates span a wide range of spectral types, from as early
as G1 to as late as M6.5.  Given ages of 1-10 Myr based on PMS
evolutionary tracks such as Siess et al. (2000) and
Baraffe et al. (1998), these spectral types correspond to a mass
range of roughly 2.5-0.1 $\msun$.  One object in IC 348 is a likely
brown dwarf ($M=0.06 \; \msun$ according to the Baraffe tracks),
and was presented in more detail in Muzerolle et al. (2006).
The distribution of spectral types is shown in Figure~\ref{spt}.
For the sample as a whole, the number of transition objects rises
from G1 to $\sim$M5, and then drops off sharply thereafter
(probably reflecting the completeness limit of the {\it Spitzer} observations).
For comparison, the stellar IMF from Kroupa (2001) is also shown
in Figure~\ref{spt}; stellar mass was converted into spectral type
using the Siess et al. (2000) pre-main sequence tracks and the effective
temperature scale given by Kenyon \& Hartmann (1995).
There appears to be a deficit of M-type stars with classical transition disks
compared to the IMF, although the small sample size makes a
quantitative statistical comparison difficult.
We conclude that there is some evidence that the kind of inner disk
clearing behavior traced by our sample may have a stellar mass dependence,
where objects with $M \lesssim 0.3 \msun$ are less likely
to exhibit a transition SED.

There is a hint of a spectral type dependence in our sample as a function
of age.  The three youngest regions lack any transition disks around stars
later than M1, while the transition sample in the four older regions
is dominated by mid-M spectral types.  Despite the small statistics,
we believe this may be a robust result since the parent population of spectral
types of stars around which we can detect disk emission out to 24 \microns
for each cluster peaks strongly at about M4-M5.
An apparent counter-example to this possible age dependence is
the 10 Myr-old TW Hydrae association, which has three transition objects
all with spectral types earlier than M4 (Table~\ref{known}).
However, the median spectral type of the total sample of $\sim$20 confirmed
stellar members (Webb et al. 1999; Sterzik et al. 1999; Zuckerman et al. 2001)
is M1, and there is only one known stellar member later than M3.5,
reflecting either a bias from the X-ray and proper motion techniques used
to identify members or an anomalous initial mass function (Moraux et al. 2007).
In any case, we cannot draw definitive conclusions without a larger sample.
However, we do note that studies of the overall disk frequency show a similar
mass-age dependence where more massive stars tend to lose their disks earlier
(Lada et al. 2006; Carpenter et al. 2006; Kennedy \& Kenyon 2009).

We also looked for signs of accretion activity in these disks using
spectroscopic indicators such as emission lines.
Quiescent (``passive") or accreting disks can be crudely distinguished
on the basis of H$\alpha$ emission equivalent width.  Adopting the
spectral type-dependent criteria of White \& Basri (2003),
we can separate chromospheric from accretion emission for all objects
with H$\alpha$ measurements (Table~\ref{holes}).
We find that 9/17 classical transition disks in our sample show evidence for
active accretion.  This may be a lower
limit since much of the H$\alpha$ data are based on low-resolution spectra;
some weak accretors can exhibit line equivalent widths below the adopted
threshold yet still show signposts of gas infall in velocity-resolved
line profiles (e.g. Flaherty \& Muzerolle 2008).  High spectral resolution
observations for some objects have been reported in the literature,
and we mark them in Table~\ref{holes}.  None of the accretor classifications
change as a result, but we point out that all but one of the ``non-accretors"
have yet to be observed at high resolution and remain untested.
High-resolution spectroscopy of H$\alpha$ for the full sample
is required to obtain the best constraints on any residual accretion activity.
Even with velocity-resolved data, all quoted accretor fractions should
be considered lower limits, since very weak accretion may not produce any
observable diagnostic.
In any case, for accretion to occur, gas must be present in these disks
very close to the star ($< 0.1$ AU).  Thus, the appearance of accretion
signatures in over half of the transition objects with spectroscopic
information indicates that the presence of gas in the inner optically thin
regions of transition disks is surprisingly common.

\begin{figure}
\plotone{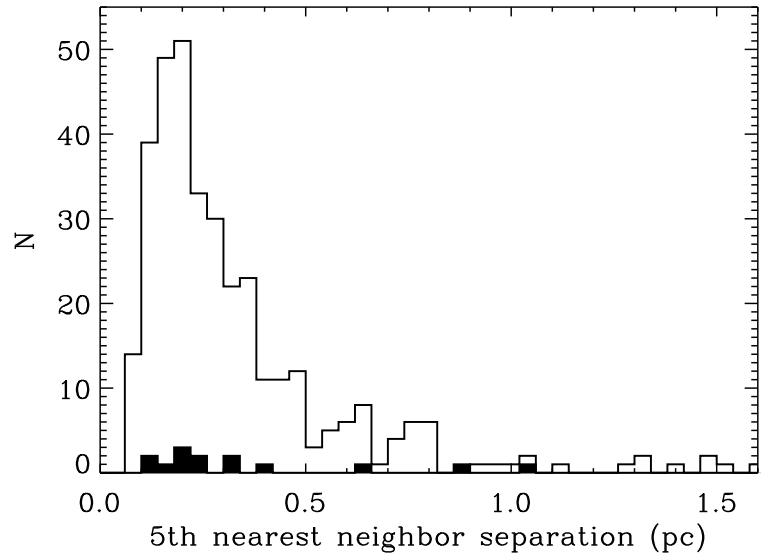}
\caption{Fifth-nearest neighbor distribution for the classical transition disks
(solid histogram) and all other disks detected out to 24 \microns
(open histogram) for the combined sample in L1688, NGC 1333, NGC 2068/2071,
and IC 348.
\label{nndist}}
\end{figure}

To test for any effect of environment on the clearing of inner disk holes,
we compared the spatial distribution of
our transition disk sample.  We did not include the three older associations,
since they are more diffuse and the {\it Spitzer} coverage of their total
membership is likely incomplete.  The four younger clusters are all primarily
low-mass star forming regions (no O stars, only one or two B stars) and
have similar stellar densities.  Figure~\ref{nndist} shows the distribution
of fifth-nearest neighbors for the transition disks versus all other disks
for the combined cluster sample.  As advocated by
Gutermuth et al. (2005), the fifth-nearest neighbor statistic
should be a reasonable compromise,
weeding out statistical outliers and wide binaries while
still accounting for non-spherical substructure common in young clusters.
Although the small numbers prevent a reliable quantitative statistical test,
there is no obvious difference between the two distributions.
This suggests that whatever process controls the transition phase
is not directly linked to either local stellar density or
radiation environment, at least in the rather limited range probed
by the clusters considered here.

Stellar multiplicity is another important property that can have a
significant effect on inner disk clearing.  Close companions can clear
a hole in the circumbinary disk, as in the case of CoKu Tau/4, while wide
companions may have an indirect influence on circumprimary disk evolution.
We unfortunately have little information concerning the multiplicity
of the transition sample.  None of the objects have resolved companions
in either the 2MASS or IRAC images, which places separation limits of
roughly $a<320-800$ AU depending on the distance.  This is not very stringent
since the majority of known young binary systems have smaller separations.
Only two of the sample regions, IC 348 and $\eta$ Cha, have been subject to
extensive high spatial resolution binary searches.  In the case of IC 348,
Duch\^ene et al. (1999) observed transition objects LRLL 21,
67, and 133, and found no companions with mass ratio $q \gtrsim 0.1$ and
separation $a \gtrsim 50$ AU; none of the other transition objects
were observed.  For the $\eta$ Cha association, Brandeker et al. (2006)
found no companion with mass ratio $q \gtrsim 0.1$ and separation
$a \gtrsim 20$ AU around the transition object RECX 5.
Finally, a recent VLBA study of DoAr 21 in L 1688 detected a stellar
companion with separation of roughly 1.5 AU (Loinard et al. 2008).

\subsubsection{Age trend}

There is some indication of a possible trend with age, in that the three
youngest clusters in our sample have a significantly lower classical transition
disk fraction ($f_{hole}$) than the other four regions.  Some of this trend
might be a result of the overall decline in total disk fraction ($f_{disk}$)
with time (see the upper panel of Figure~\ref{holefrac}), since we have
calculated $f_{hole}$ relative to the disked rather than total stellar
population of each cluster.
We can estimate $N_{hole}/N_*$ if $f_{disk}$ is known and assuming that
our transition disk sample is complete.  The disk fraction has been estimated
from {\it Spitzer} data for most of the regions in our sample: NGC 1333
(83\%; Gutermuth et al. 2008), NGC 2068/2071 (75\%; Flaherty \& Muzerolle 2008),
IC 348 (47\%; Lada et al. 2006), $\eta$ Cha (40\%; Megeath et al. 2005),
and Orion OB1a and b (6\% and 13\%, respectively; Hern\'andez et al. 2007).
We then derive
$N_{hole}/N_* \sim f_{trans} \times f_{disk} \sim 1.2^{+3.0}_{-1.0}$,
1.5$^{+1.4}_{-0.8}$, 5.6$^{+2.4}_{-2.2}$, 1.0$^{+2.0}_{-0.8}$,
6.6$^{+11}_{-6.4}$, and 1.0$^{+1.8}_{-1.0}$ \% for the 7 regions
in order of increasing age.  These percentages are clearly not equal,
although the age trend is not well-defined or consistent.
Further observations of other regions in the crucial 3-10 Myr age range
are needed to improve the statistics.
We do emphasize that there are biases that can effect these estimates.
We have already discussed detection limits.  Also, assigning ages to
young stars is fraught with complications, being model dependent and
subject to significant biases from variability and accretion effects
and uncertain age spreads (at least within a few Myr) among individual objects.
We have chosen to make qualitative
comparisons based on mean ages for each region rather than adopting ages
for individual objects.  It is possible the transition objects are
systematically older than the mean age of the clusters, but we find
no statistically significant evidence of a such a bias in the HR diagrams
of these regions.

If the trend with mean age is real, it may reveal important clues
to the mechanisms that create the inner disk clearing and the eventual
dissipation of the entire disk.  We looked at this in more detail
by constructing a parametric model of disk dissipation using Monte Carlo
techniques.  The observed decline in $f_{disk}$ with time can be used
to trace the overall disk dissipation rate.  We have collected all
estimates of $f_{disk}$ from {\it Spitzer} observations of young clusters
and associations, and show them in the upper panel of Figure~\ref{holefrac}.
All but one of these estimates is based on the presence of excess
emission at $\lambda \leq 8$ \microns from IRAC imaging data; the exception
is the TW Hydrae association, for which we estimated $f_{disk}$
from the MIPS data in Low et al. (2006), selecting all objects with
$L_{IR}/L_* \gtrsim 0.1$ indicative of optically thick disks.
Most of the regions are near enough ($d \lesssim 500$ pc) for
the disk inventories to be complete down to or below the substellar limit.
We see a clear decline with age with relatively small scatter --
the only significant outliers are Tr 37 at $t \sim 4$ Myr and $\eta$ Cha
at $t \sim 7$ Myr; the former likely contains a mixed stellar population
from separate star formation episodes (Sicilia-Aguilar et al. 2004),
while the latter has only 15 known members and may have been subject to
significant dynamical evolution.
The drop in $f_{disk}$ is initially quite rapid, down to $\sim 40$\%
by 3 Myr, then appears to tail off into a more gradual decline
at older ages.

Based on these observations, we adopted a two-component
parameterization to describe the disk dissipation rate for our model.
The exact functional form, and a physical motivation for it, is
not important here; we do note that a single exponential decay
cannot match the observed tail at 5-10 Myr.
The resulting simulated disk fraction at each time step from $t=0$
to 10 Myr, assuming an initial disk fraction of 100\%, was calculated
for an ensemble of $10^5$ stars.  We adopted a gaussian age
distribution with 1$\sigma \sim 1$ Myr centered at $t$ for the ensemble
to reflect the typical small age spread in young clusters.
Finally, we assumed that as each disk dissipates, it immediately forms
an AU-scale inner dust hole and remains in that state for
a time $t_{hole}$, during which the disk exhibits our defined transition
characteristics and after which all disk material detectable out
to 24 $\mu$m has completely dissipated.
We then varied $t_{hole}$ to best match the observed transition disk
fraction (lower panel of Fig.~\ref{holefrac}).
A value of $t_{hole}=$ 0.1 Myr, similar to previous estimates derived
from simple statistical arguments, results in an order-of-magnitude
agreement.  The actual number of transition disks in the simulated ensemble
is constant with time within each of the two decay intervals
($0-3$, $3-10$ Myr),
since we assume a constant decay slope.  The rise in the model $f_{hole}$
from 0-2 and 5-10 Myr is a consequence of the decrease in the total disk
fraction, while the dip at 2-5 Myr is produced by the ``elbow" between
the two decay intervals where the adopted decay slope changes.
However, the model does not reproduce the data in detail; it slightly
overpredicts $f_{hole}$ at 1-3 Myr, and underpredicts $f_{hole}$
at 3-10 Myr.

Better agreement with the data might a be accomplished by adopting an
age-dependent $t_{hole}$.  Figure~\ref{holefrac} shows a second model
with $t_{hole}=$ 1 Myr (the decay rates were adjusted in this case
in order to match the observed $f_{disk}$, otherwise the model would
vastly overpredict the disk fraction at 1-3 Myr); this model provides
a good match to observations at 5-10 Myr.
Alternatively, it is possible that only a fraction of the disks that
dissipate at $t \lesssim 3$ Myr actually go through an identifiable
transition phase, while the rest dissipate much more quickly through
other means (e.g. disruption from encounters or binary interactions).
This dichotomy may also be related to the spectral type dependence
indicated in Figure~\ref{spt}, which our simple model does not take into
account.  A more sophisticated model awaits better characterization
of mass-dependent disk dissipation rates.  We can estimate this in
a very rough way by considering two spectral type bins: K0-M1, which
encompasses all of the young (1-2 Myr) transition objects, and M1-M5,
which encompasses most of the older (3-10 Myr) transition objects.
The fraction of all cluster stars in each bin, assuming a typical IMF,
is roughly 30\% and 70\%, respectively.
Since the young transition disks are only in the early-type bin,
while all young disks span the full range of spectral types,
we can estimate an ``IMF-weighted" transition disk fraction of
$0.02/0.3 = 7\%$ at 1-2 Myr.  This is closer to the 7-17\% range at 3-10 Myr,
although an absolute age dependence cannot be confidently ruled out.
In any case, our results may indicate
that different disk dissipation mechanisms may operate at different times.
Specific physically-motivated models are needed to explore this in more detail.
In addition, more data in the 5-10 Myr age range are needed to better
constrain the age trend; unfortunately, statistics will be hard to come by
since the overall disk fractions are so small at older ages.





\subsection{Warm and weak excess transition objects}

The remaining sources flagged as possible transition disks by their
spectral slopes but rejected based on other criteria are nevertheless
interesting since they still exhibit some signs of significant evolution.
The warm excess sources show some continuum excess emission
at $\lambda=3-8$ \microns that, while in many cases may be optically
thick, is still significantly below ``typical" disk emission at
those wavelengths.  Meanwhile, the spectral
slope at 8-24 \microns increases steeply in all cases.  These SED dips may
indicate a substantial lack of dust at disk radii somewhere outside of
the dust sublimation radius but within about 1-20 AU.  Similar behavior
has been recently reported in a number of disks around nearby young
solar-type stars (Brown et al. 2007) and 3 stars in Taurus
(Espaillat et al 2007).  Radiative transfer models of these objects
are able to reproduce their SEDs by including substantial disk gaps with
inner radii $R \sim 0.1 - 2$ AU and outer radii $R \sim 5 -50$ AU.
Hence, these objects may be precursors to an inner hole phase
(so-called ``pre-transitional" disks; Espaillat et al. 2007).
Interestingly, the fraction of measurable accretors in this group
(5/6; Table~\ref{marginal}) is larger than in the classical and weak excess
transition categories (though admittedly the sample size is very
small).  Two of these objects
have H$\alpha$ equivalent widths below the adopted threshold,
but one (FM 326) does show a weak accretion component
in the line profile at high spectral resolution (Flaherty \& Muzerolle 2008).
Note that the single non-accretor has
not been observed at high spectral resolution, so it is possible that
all six of these sources are still accreting.

The weak excess sources have mid-IR fluxes that suggest marginally
optically thick if not optically thin disk material at the radii probed
by that wavelength.  It is possible that some of these SEDs might rise
at $\lambda > 24$ \micron, indicating very large inner holes.
One such object in the nearby Lupus star forming region was discovered
by Padgett et al. (2006), although its fractional disk luminosity
of $\sim 8 \times 10^{-4}$ suggests its outer disk may be optically thin.
One of our sources, FM 458, is coincident with a bright MIPS 70 \microns
(Flaherty \& Muzerolle 2008) and SCUBA sub-mm source (Mitchell et al. 2001).
However, the long-wavelength luminosity suggests a very large mass of
circumstellar material, and the source may be a chance superposition of a
T Tauri star with a Class 0/I object.
In any case, the {\it Spitzer} data generally lacks sufficient sensitivity
at 70 \microns (because of the greater distances and much stronger
background emission) for better statistics on any disks with
extremely large inner holes.

\begin{figure}
\plotone{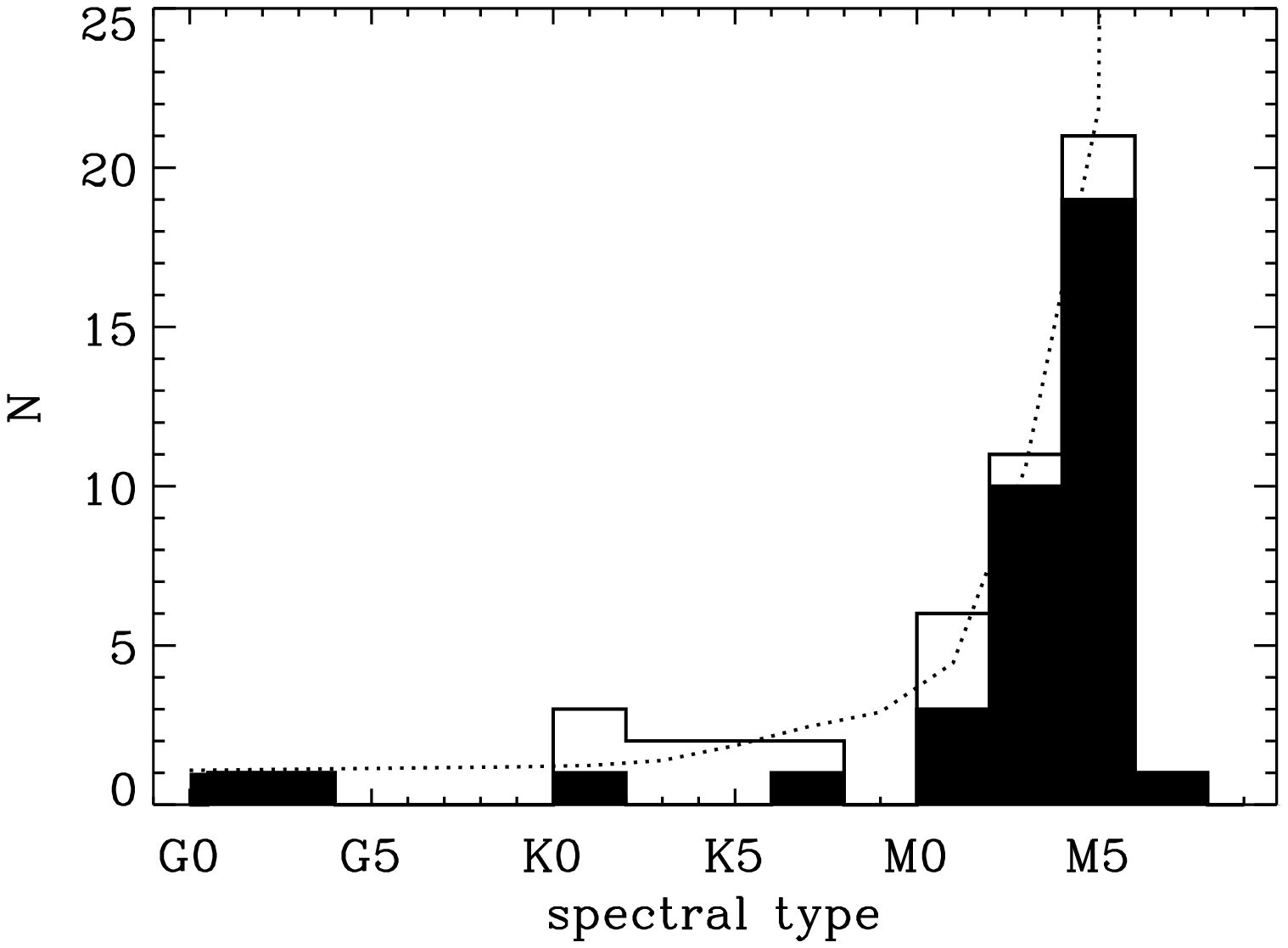}
\caption{Same as in Figure~\ref{spt}, with the warm and weak excess
transition objects added in.
\label{spt_add}}
\end{figure}

There are two characteristics of note for the weak excess sample.
First, their accretion activity appears to be significantly reduced if
not absent.  Only 3/20 objects exhibit H$\alpha$ equivalent widths above
the accretion threshold.  Again, we could be missing some weak accretors
since high spectral resolution observations are lacking for most of the sample.
Broad and/or asymmetric line profiles suggestive of weak accretion
have been observed in three objects, FM 458 and FM 998 (Flaherty \& Muzerolle
2008) and RECX 9 (Lawson et al. 2004), which brings the accretor fraction
to 6/20, or $\sim 30$\%.  Nevertheless, in comparison with the classical
and warm excess transition objects, the relative H$\alpha$ strengths indicate
that the weak excess sources are more weakly accreting on average.
Second, these sources are strongly associated with lower mass stars;
20/25 have spectral types later than M0, and the distribution of types
rises sharply to our completeness limit of M4-M5.  This behavior may be
analogous to the passive disks reported by  McCabe et al. (2006),
which showed weak/absent accretion and appeared only around M stars.
Figure~\ref{spt_add} shows the spectral type distribution of the combined
transition sample.  In marked contrast to the classical transition sample,
the combined distribution
strongly peaks at later spectral types, and is much more consistent with
the stellar IMF.  Since the weak excess objects with mid-M types
are more likely to be affected by detection biases because of their
weaker 24 \microns fluxes, the combined distribution
further bolsters the argument that the classical transition objects may be
underrepresented among mid- to late-M stars.

The trend of increasing transition disk fraction with
age is greatly enhanced if the warm and weak excess
sources are included.  In that case, we would get
1.5\% (1/66), 2.7\% (2/74), 6.6\% (10/152), 26\% (26/99), 29\% (4/14),
67\% (4/6), and 67\% (4/6) for the 7 regions in our study in rough age order
(see Fig.~\ref{holefrac}).  These statistics are less robust
since the weak excess sample is likely not complete, particularly for
the more distant young clusters with higher background emission.
Nevertheless, there is a clear trend of increasing frequency of
highly evolved disks as a function of age, as has been seen by prior
studies looking at spectral slope evolution.

\section{Discussion}

Understanding the properties of circumstellar disks with significant dust
clearing in their inner regions may have profound implications for
planet formation processes.  If indeed these transition disks are showing us
the end-stages of protoplanetary disks, their statistics are indicative
of disk dissipation mechanisms and timescales, which can then be used to
constrain models of planetesimal growth and/or giant planet accretion.
Some of the objects in our sample have ages of 1 Myr or less, which implies
that the disk clearing mechanism(s) must operate very quickly.
Our simple parametric model of disk dissipation indicates a characteristic
``transition" time in the range $0.1 - 1\times 10^6$ years, possibly
dependent on the stellar age and/or mass.
There have been competing claims in the recent literature regarding
the transition timescale, with some advocating for a rapid transition of
order $10^5$ years (e.g., Furlan et al. 2009) and others arguing for
a much longer timescale comparable to the stellar age
(e.g., Sicilia-Aguilar et al. 2008; Currie et al. 2009ab).
Much of the discrepancy may lie in the wide range of adopted definitions
of what constitutes a transitional disk; the studies that derive longer
transition times tend to use much less restrictive selection criteria.
Our results support a slower transition for disks only around stars older
than $\sim 3$ Myr.  As we mentioned in section 3.3.2, it is difficult
to directly compare statistics for regions of different ages because
they also appear to probe different stellar mass regimes.
This uncertainty can only be addressed once the dependence of disk
evolution on stellar mass has been better-delineated.
All of the transition times were estimated assuming that every disk
goes through the same inner hole phase.  However, our observation of
a possible deficit of classical transition disks around late-type stars,
as well as the apparent age dependence of the transition disk frequency,
appears to contradict this assumption.  If classical transition disks
do not represent a universal phase of disk evolution, then the clearing
timescale may be longer than our estimates.

The presence of accretion, and hence gas within the dust hole,
in a majority of the classical transition objects presents challenges
to understanding how the inner disk is cleared of small grains
while the outer disk (given the significant excesses at
$\lambda > 10$ \micron) is not.
One possibility is that significant grain growth to meter-size or larger
bodies has occurred in the inner disk, as has been suggested by
models of grain coagulation (Weidenschilling 1997; Dullemond \& Dominik 2005).
However, some other process is then needed
to decouple the outer disk so that accretion from there cannot
replenish the supply of small grains to the inner disk.  In this scenario,
the inner disk should then drain onto the star in a viscous time,
which for typical T Tauri parameters is of order a few $10^5$ years,
similar to some of the clearing timescales estimated above.
Several gap-creating mechanisms have been proposed for T Tauri disks,
including photoevaporative mass loss induced by UV radiation from
the central star (e.g., Clarke et al. 2001) and the dynamical influence
of a giant planet (e.g., Bryden et al. 1999).  Najita et al. (2007)
review these mechanisms and suggest that no single one operates in all cases.
Our results support this view, particularly if the hole clearing timescale
or incidence really does vary with stellar age and mass.
The weak excess transition objects may represent a stronger case
for the photoevaporation mechanism: the low or absent accretion activity
and low disk masses as implied by the small fractional infrared excess
are both properties predicted by the models at the time of inner hole
formation (see also Cieza et al. 2008).  However, the large number
of these objects, especially at older ages, may not be consistent with
photoevaporation model predictions that the outer disk should dissipate
very rapidly following inner hole formation (less than $10^5$ years;
Alexander et al. 2006).  This may be resolved if photoevaporation is
less efficient for M stars; current models have not yet fully explored
the effect of stellar mass.

The inner disk clearing we see in the accreting transition objects
(if not all of them) is most likely due to an advanced
stage of dust evolution, either from growing planetesimals or possibly
newly formed planets.
Such processes occur across a large range of ages,
from the youngest $\sim 1$ Myr embedded star forming regions to some
of the oldest known classical T Tauri stars such as the 10 Myr-old
TW Hya, with some evidence for an increasing frequency with age.
The cause of such disparate onset times for inner disk clearing
is a key question that remains unanswered.  Initial conditions, such as
the initial disk size and mass, may play a role.
For example, disks of the same mass that are born with different initial
sizes will viscously evolve at different rates, may form planets at
different times, and may have different susceptibility to photoevaporation.
All of these properties can influence the onset time of inner clearing
and the amount of time required for the outer disk to dissipate.
More massive stars
may be born with more massive disks, which might have the best chance
for forming giant planets.  Such a scenario could explain the young transition
disks, which are preferentially associated with more massive stars.
It may also explain the deficit of classical transition disks around
the lowest mass
stars, if in fact photoevaporation turns out not to be an efficient gas
removal mechanism.  Alexander \& Armitage (2009)
have explored some of these parameters by calculating disk evolution models
including the effects of giant planet formation, migration, and
photoevaporation.  Interestingly, the transitional disk fraction
they predict as a function
of stellar age is qualitatively similar to our observed trend (although
their values at 5-10 Myr are somewhat higher).  However, they used
a solar-mass star for all models, so the stellar mass dependence
remains to be considered.

We have shown that most classical transition disks exhibit trace amounts of
excess emission at 5.8 \micron, indicating that the inner holes are not
totally devoid of small grains.  This is consistent with models of
{\it Spitzer}-IRS spectra of transition disks published to date, which can only
explain the observed 10 \microns silicate emission in most cases by placing
optically thin dust inside the hole (e.g. Calvet et al. 2005).
The origin of this dust is unclear.  It is possible that in objects that
are still accreting, some dust is entrained in the accretion flow passing
from the outer disk into the hole; numerical simulations by Rice et al. (2006)
of a disk gap cleared by a giant planet show that pressure
gradients at the outer edge of the gap can filter dust particles,
allowing only small grains to pass through the gap along with the gas
accretion flow.
However, many of the non-accreting transitional systems also show
hot dust excess, although some of these may be accreting at rates too low
to produce observable diagnostics.  Some objects, particularly those with
larger $f_{5.8}$, may still have optically thick material near
the dust sublimation
radius.  In that case, the emission may have been reduced from that of
typical disks because of a lower scale height brought about by dust grain
growth and settling.  An asymmetric inner disk geometry such as
a warp might also produce lower excess emission at certain orientations;
this may manifest itself through mid-infrared variability, which has been
recently discovered in the transitional disk LRLL 31 (Muzerolle et al. 2009).
Finally, there may also be {\it in situ} dust production
via colliding planetesimals, as Eisner et al. (2006) proposed for TW Hya,
possibly tracing active terrestrial planet formation.

One obvious question is whether the three classes of evolved disks
identified here are connected.
We would argue that the classical and warm excess transition objects
do likely represent an evolutionary sequence.
The warm excess sources are akin to the pre-transitional disks
described by Espaillat et al. (2007), and are most likely produced as
a result of embedded giant planets carving out large gaps in the disks
(Espaillat et al. 2008).  As the planet continues to grow via accretion
from the outer disk, it should eventually stifle accretion through the gap
and onto an optically thick inner disk, which may render the interior
completely optically thin and result in a classical transition disk.
However, the link between these and the weak excess disks is much more
dubious.  The sheer numbers of the weak excess objects,
particularly around older stars, are too great to simply be the result of
further evolution of the outer disks of classical transition objects.
The spectral type distributions are also distinctly different.
Other studies have suggested that there may be separate disk
evolutionary pathways (e.g. Lada et al. 2006; Cieza et al. 2007)
of which the different disk classes may be representative.
However, the weak excess category is less well-defined,
and very likely contains a mixture of different disk types.
The identification of true inner holes is not as robust for the weak
excess disks around lower mass stars, as recently pointed out by
Ercolano et al. (2009).  They may instead represent evolution without
a cleared inner hole phase.  Similar objects have been dubbed ``homologously
depleted" (Wood et al. 2002; Currie et al. 2009b), suggesting that
the dust opacity has been reduced through grain growth and mass loss
across a wide range of disk radii.  We would like
to point out that dust grain growth and settling acts to shrink the
effective range of disk radii probed at a given wavelength; the models of
D'Alessio et al. (2006) show that the majority of dust emission at 25 \microns
is located within $\sim 1-2$ AU in a highly settled disk.  Thus, radial dust
evolution in these disks cannot be definitively ruled out without
measurements at far-infrared wavelengths.
It is also very possible that some of the weak excess transition disks
may in fact be recently-formed debris disks; further study of their gas
content is required before this distinction can be made.

The recent discovery that the CoKu Tau/4 disk is circumbinary
(Ireland \& Kraus 2008) has fueled speculation as to whether all
transitional disks are in fact the result of disk clearing by a
stellar companion, and indeed whether such objects are truly transitional.
We argue that regardless of hole formation mechanism, all objects
showing these SED characteristics are likely to be in transition.  There are
examples of close binaries whose SEDs betray no evidence for significant
inner holes (DF Tau, separation $\sim 13$ AU, and GW Ori, separation
$\sim 1$ AU, to name a few), which suggests that disks around close
binary systems can and do evolve in a similar fashion.

Unfortunately, we lack binary statistics for
our sample, particularly at close separations ($< 10$ AU).
We can indirectly assess the importance of binaries by comparing
the expected fraction of binaries with our observed
transition disk statistics as a function of stellar age and mass.
From the multiplicity study by Sterzik \& Durisen (2004), we estimate
a total multiplicity fraction of about 50\% for solar-mass stars,
and roughly 20\% at $0.2 \; \msun$.  Since our transition disk selection
is sensitive to maximum inner disk hole sizes of roughly 40 AU for
solar type stars and 4 AU for low mass stars, any stellar companions
must be within this range in order to produce the disk clearing.
According to Sterzik \& Durisen (2004),
about 50\% of the $1 \; \msun$ binaries have separations $<40$ AU
while about 30\% of the $0.2 \; \msun$ binaries will
have separations within 4 AU, leading to combined fractions of 25\% and 6\%,
respectively.  These estimates are inconsistent with the transition disk
statistics in both magnitude and trend with mass, thus we suggest that
stellar companions are unlikely to be a significant contributor.
Nevertheless, sensitive searches for close stellar companions, both with
AO imaging and radial velocity monitoring, need to be done before
firm conclusions can be made.

\begin{figure}
\plotone{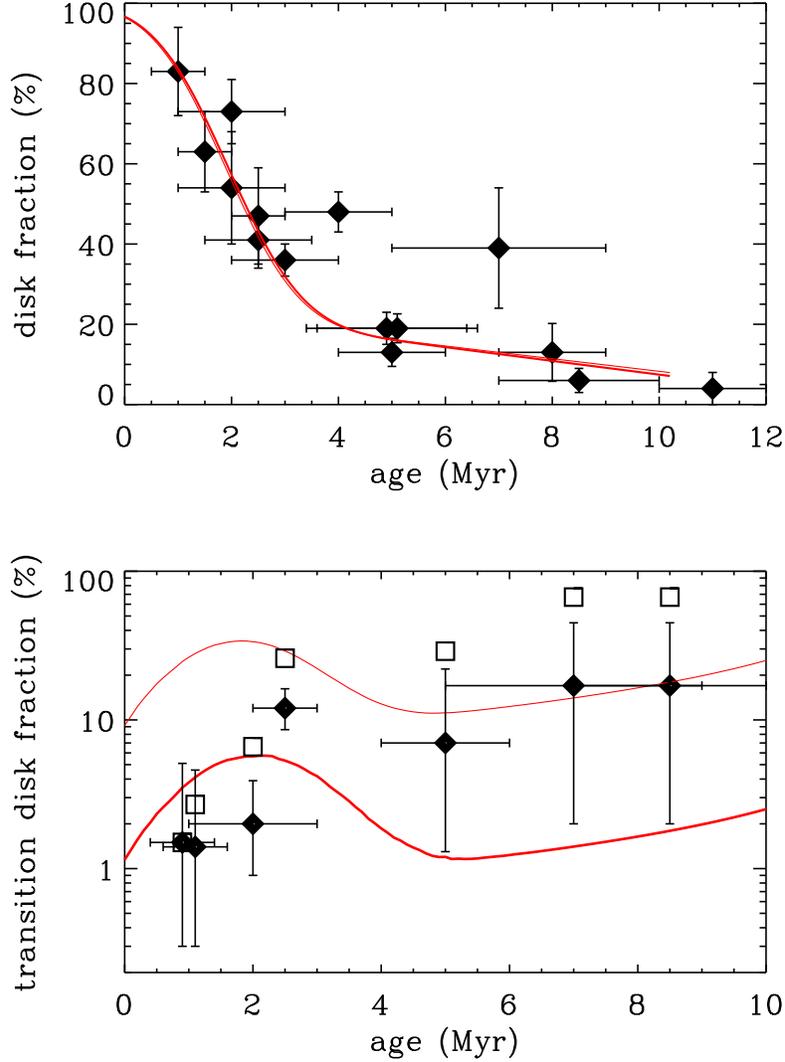}
\caption{{\it Top panel:} The fraction of all members of a given young
cluster or association that exhibit infrared excess indicative of
protoplanetary disks (data from Gutermuth et al. 2008; Furlan et al. 2006;
Flaherty \& Muzerolle 2008; Lada et al. 2006; Balog et al. 2006;
Gutermuth et al. 2004; Sicilia-Aguilar et al. 2006;
Hern\'andez et al. 2007ab; Carpenter et al. 2006; Dahm \& Hillenbrand 2007;
Megeath et al. 2005; Low et al. 2006).
The points near $t=5$ Myr, $f_d=20$\% (representing Upper Sco and NGC 2362)
have been shifted slightly in age for clarity.  The solid line shows
our adopted two-component parametric model for disk dissipation with
two different transition timescales: 0.1 Myr (thick) and 1 Myr (thin).
{\it Bottom panel:} The percentage of protoplanetary disks in a given region
that exhibit SEDs consistent with classical transition disks (diamonds).
The open squares represent the percentages including the warm and weak
excess classes.
The solid lines show the predicted values from the dissipation model
for the 0.1 and 1 Myr transition timescales.
\label{holefrac}}
\end{figure}

Finally, the possibility of different disk decay rates at
different stellar ages, as indicated in Figure~\ref{holefrac}, is very
intriguing in its own right.  It may be a further indicator that
different disk dissipation mechanisms operate preferentially at
different times.  The stellar mass dependence, where disks around less
massive stars tend to last longer, may also be involved.  For example,
if either photoevaporation or planet formation is less effective in
mid- to late-M stars, their disks
may take longer to completely erode.  Alternatively, stellar multiplicity
may have an effect; evidence suggests that disks tend to disappear more
quickly around binaries (e.g. Bouwman et al. 2006), and these may represent
the faster decay rate at younger ages.  However, counter-examples
(long-lived disks around binaries, short-lived disks around single stars)
are plentiful.  More work needs to be done to fully elucidate the effects
of these parameters, as well as others such as environment and initial
conditions.

\acknowledgements

This work is based on observations made with the {\it{Spitzer Space
Telescope}}, which is operated by the Jet Propulsion Laboratory,
California Institute of Technology under NASA contract 1407.
We also used data from the Two Micron All Sky Survey (2MASS), a joint
project of the University of Massachusetts and the Infrared Processing
and Analysis Center (IPAC)/California Institute of Technology,
funded by NASA and the National Science Foundation.
We thank an anonymous referee for constructive and insightful suggestions.
Support for this work was provided by NASA through Contract Number
960785 issued by JPL/Caltech.

\begin{deluxetable}{lccr}
\tabletypesize{\small}
\tablewidth{0pt}
\tablecaption{Sample of young stellar clusters/associations\label{clusters}}
\tablehead{
\colhead{ID} &
\colhead{$t$ (Myr)} &
\colhead{$d$ (pc)} &
\colhead{transition disk fraction}}
\startdata
NGC 1333$^a$ & $<1^i$ & 320 & 1/66 (1.5$^{+3.6}_{-1.2} \,$\%)\\
L 1688$^b$ & $<1^j$ & 140 & 1/74 (1.4$^{+3.2}_{-1.1} \,$\%)\\
NGC 2068/2071$^c$ & 1-3$^k$ & 400 & 3/152 (2.0$^{+1.9}_{-1.1} \,$\%)\\
IC 348$^{d,e}$ & 2-3$^l$ & 315 & 12/99 (12$^{+4.2}_{-3.4} \,$\%)\\
OB1b$^f$ & 4-6$^k$ & 400 & 1/14 (7.1$^{+15}_{-5.7} \,$\%)\\
$\eta$ Cha$^{g,h}$ & 5-9$^n$ & 100 & 1/6 (17$^{+28}_{-15} \,$\%)\\
OB1a/25 Ori$^f$ & 7-10$^m$ & 330 & 1/6 (17$^{+28}_{-15} \,$\%)\\
\enddata
\tablecomments{{\it Spitzer} data from: $^a$Gutermuth et al. (2008);
$^b$Allen et al. in preparation; $^c$Flaherty \& Muzerolle (2008);
$^d$Lada et al. (2006); $^e$Muench et al. (2007);
$^f$Hern\'andez et al. (2007); $^g$Megeath et al. (2005);
$^h$Gautier et al. (2008).  Age estimates
are taken from the literature: $^i$Wilking et al. (2004);
$^j$Luhman et al. (2001); $^k$Flaherty \& Muzerolle (2008);
$^l$Luhman et al. (2003); $^m$Brice\~no et al. (2005);
$^n$Luhman \& Steeghs (2004).}
\end{deluxetable}

\begin{deluxetable}{lcccccccc}
\tabletypesize{\small}
\tablewidth{0pt}
\tablecaption{Previously identified classical transition disks\label{known}}
\tablehead{
\colhead{ID} &
\colhead{$\alpha$(J2000)} &
\colhead{$\delta$(J2000)} &
\colhead{SpT} &
\colhead{A$_V$} &
\colhead{$f_{5.8}$} &
\colhead{W(H$\alpha$)} &
\colhead{accretor?}}
\startdata
DM Tau$^a$ & 04:33:48.73 & +18:10:10.0 & M1 & 0.5 & \nodata & -139 & y\\
CoKu Tau/4$^b$ & 04:41:16.81 & +28:40:00.0  & M1 & 3.0 & 1.0 & -1.8 & n\\
GM Aur$^{a,c}$ & 04:55:10.98 & +30:21:59.4 & K5 & 1.2 & 1.5 & -97 & y*\\
CVSO 224$^d$ & 05:25:46.74 & +01:43:30.4 & M3 & 0.2 & 1.3 & -20.3 & y*\\
TW Hya$^{c,e}$ & 11:01:51.01 & -34:42:17.0 & K7 & 0.0 & 1.3 & -220 & y*\\
Hen 3-600A$^e$ & 11:10:27.88 & -37:31:52.0 & M3 & 0.0 & \nodata & -22 & y*\\
HD 98800$^f$ & 11:22:05.29 & -24:46:39.8 & K5 & 0.0 & \nodata & 0 & n*\\
CS Cha$^g$ & 11:02:25.12 & -77:33:36.0 & K6 & 0.8 & \nodata & -20 & y*\\
\enddata
\tablecomments{Based on {\it Spitzer} data from: $^a$Calvet et al. 2005;
$^b$D'Alessio et al. 2005; $^c$Hartmann et al. 2005; $^d$Espaillat et al. 2008b;
$^e$Uchida et al. 2004; $^f$Furlan et al. 2007; $^g$Espaillat et al. 2007.
Accretor status is based on the H$\alpha$ equivalent width, as explained in
the text, except for those objects marked with $*$, where it is based on
published high spectral resolution H$\alpha$ profiles.}
\end{deluxetable}

\clearpage

\begin{deluxetable}{lccccccc}
\tabletypesize{\small}
\tablewidth{0pt}
\tablecaption{Classical transition disk candidates\label{holes}}
\tablehead{
\colhead{ID} &
\colhead{$\alpha$(J2000)} &
\colhead{$\delta$(J2000)} &
\colhead{SpT} &
\colhead{A$_V$} &
\colhead{$f_{5.8}$} &
\colhead{W(H$\alpha$)} &
\colhead{accretor?}}
\startdata
DoAr 21$^a$ & 16:26:03.01 & -24:23:37.9 & K0 & 6.3 & 1.5 & -0.6 & n\\
LAL 31 & 03:29:29.26 & +31:18:34.9 & K5 & 4.5 & 1.1 & -8.4 & y\\
LRLL 21$^b$ & 03:44:56.15 & 32:09:15.5 & K0 & 5.4 & 1.5 & -4.7 & y*\\
LRLL 31$^b$ & 03:44:18.16 & 32:04:57.0 & G1 & 11.9 & 1.6 & -11 & y*\\
LRLL 67$^b$ & 03:43:44.62 & 32:08:17.9 & M0.75 & 2.4 & 1.2 & -35 & y*\\
LRLL 72$^b$ & 03:44:22.57 & 32:01:53.7 & M2.5 & 2.5 (3.5) & 1.0 & -3.5 & n\\
LRLL 97$^b$ & 03:44:25.56 & 32:06:17.0 & M2.25 & 4.6 (6.0) & 1.4 & -3.0 & n\\
LRLL 133$^b$ & 03:44:41.74 & 32:12:02.4 & M5 & 3.9 & 1.4 & \nodata & \nodata \\
LRLL 190$^b$ & 03:44:29.21 & 32:01:15.8 & M3.75 & 7.1 & 1.3 & -5.0 & n\\
LRLL 237$^b$ & 03:44:23.57 & 32:09:34.0 & M5 & 1.7 & 1.4 & -6.0 & n\\
LRLL 297$^b$ & 03:44:33.21 & 32:12:57.5 & M4.5 & 7.8 (8.4) & 1.4 & \nodata & \nodata \\
LRLL 301$^b$ & 03:44:22.70 & 32:01:42.4 & M4.75 & 6.2 & 1.3 & -40 & y\\
LRLL 316$^b$ & 03:44:57.73 & 32:07:41.9 & M6.5 & 1.2 & 1.0 & -4.0 & n \\
LRLL 1679$^c$ & 03:44:52.06 & 31:58:25.2 & M3.5 & 5.7 & 1.5 & \nodata & \nodata \\
FM 177$^d$ & 05:45:41.95 & -00:12:05.3 & K4 & 1.8 & 0.9 & -1.2 & n*\\
FM 281$^d$ & 05:45:53.11 & -00:13:24.9 & M1 & 1.1 & 1.1 & -18.1 & y*\\
FM 856$^d$ & 05:46:44.85 & +00:16:59.7 & M1 & 4.2 & 1.4 & -14.0 & y*\\
CVSO 95$^e$ & 05:31:39.00 & -01:27:46.0 & M5 & 0.0 & 1.1 & -8.9 & n\\
CVSO 224$^f$ & 05:25:46.74 & +01:43:30.4 & M3 & 0.2 & 1.3 & -20.3 & y*\\
RECX 5$^g$ & 08:42:27.10 & -78:57:47.9 & M4 & 0.0 & 1.2 & -35.0 & y*\\
\enddata
\tablecomments{Spectral types and H$\alpha$ measurements from:
$^a$Bouvier \& Appenzeller (1992); $^b$Luhman et al. (2003);
$^c$Muench et al. (2007);
$^d$Flaherty \& Muzerolle (2008); $^e$Brice\~no et al. (2005);
$^f$Brice\~no et al. (2007); $^g$Lawson et al. (2004).
If the extinction was adjusted for a better SED fit (see text),
the original value from the literature is listed in parentheses.
Accretor status is marked as in Table~\ref{known}.}
\end{deluxetable}

\clearpage

\begin{deluxetable}{lccccccc}
\tabletypesize{\small}
\tablewidth{0pt}
\tablecaption{Warm excess transition candidates\label{marginal}}
\tablehead{
\colhead{ID} &
\colhead{$\alpha$(J2000)} &
\colhead{$\delta$(J2000)} &
\colhead{SpT} &
\colhead{A$_V$} &
\colhead{$f_{5.8}$} &
\colhead{W(H$\alpha$)} &
\colhead{accretor?}}
\startdata
LRLL 58$^a$ & 03:44:38.55 & 32:08:00.7 & M1.25 & 3.7 & 1.9 & -9.0 & n\\
LRLL 110$^a$ & 03:44:37.40 & 32:12:24.3 & M2 & 4.6 (5.3) & 1.8 & -22. & y\\
LRLL 194$^a$ & 03:44:27.25 & 32:10:37.3 & M4.75 & 3.1 & 2.2 & -100. & y\\
FM 326$^b$ & 05:45:56.31 & 00:07:08.6 & K7 & 1.6 & 2.3 & -6.2 & y*\\
FM 515$^b$ & 05:46:11.86 & 00:32:25.9 & K2 & 1.3 & 3.9 & -3.8 & y*\\
FM 618$^b$ & 05:46:22.44 & -00:08:52.6 & K1 & 2.6 & 2.5 & -32 & y*\\
\enddata
\tablecomments{Spectral types and H$\alpha$ measurements from:
$^a$Luhman et al. (2003); $^b$Flaherty \& Muzerolle (2008).
If the extinction was adjusted for a better SED fit (see text),
the original value from the literature is listed in parentheses.
Accretor status is marked as in Table~\ref{known}.}
\end{deluxetable}

\clearpage

\begin{deluxetable}{lccccccc}
\tabletypesize{\small}
\tablewidth{0pt}
\tablecaption{Weak excess transition candidates\label{weak24}}
\tablehead{
\colhead{ID} &
\colhead{$\alpha$(J2000)} &
\colhead{$\delta$(J2000)} &
\colhead{SpT} &
\colhead{$A_V$} &
\colhead{$f_{5.8}$} &
\colhead{W(H$\alpha$)} &
\colhead{accretor?}}
\startdata
GY 326$^a$ & 16:27:42.70 & -24:38:50.6 & M4 & 10.0 (11.4) & 1.2 & \nodata & \nodata \\
LRLL 6$^b$ & 03:44:36.94 &  32:06:45.4 & G3 & 3.2 (3.9) & 1.5 & 0.7 & n* \\
LRLL 30$^c$ & 03:44:19.13 &  32:09:31.4 & F0 & 0.3 (1.9) & 1.0 & \nodata & \nodata \\
LRLL 68$^c$ & 03:44:28.51 & 31:59:54.1 & M3.5 & 1.8 (2.9) & 1.6 & -5.1 & n \\
LRLL 76$^c$ & 03:44:39.81 & 32:18:04.2 & M3.75 & 3.1 & 1.6 & -14 & n \\
LRLL 135$^c$ & 03:44:39.19 & 32 20:09.0 & M4.5 & 1.1 (2.0) & 1.6 & -20 & y \\
LRLL 176$^c$ & 03:45:04.63 & 32 15:01.1 & M4.25 & 4.1 & 1.0 & \nodata & \nodata \\
LRLL 182$^c$ & 03:44:18.20 & 32:09:59.3 & M4.25 & 2.5 (3.1) & 1.4 & -6.0 & n \\
LRLL 213$^c$ & 03:44:21.27 & 32:12:37.3 & M4.75 & 1.4 (2.1) & 1.7 & -6.0 & n \\
LRLL 214$^c$ & 03:44:07.51 & 32:04:08.9 & M4.75 & 1.1 (2.1) & 1.5 & -7.0 & n\\
LRLL 229$^c$ & 03:44:57.86 & 32:04:01.8 & M5.25 & 1.2 & 1.2 & -4.5 & n\\
LRLL 241$^c$ & 03:44:59.84 & 32:13:32.2 & M4.5 & 2.8 (3.2) & 1.5 & -80 & y \\
FM 458$^d$ & 05:46:07.89 & -00:11:56.9 & K3 & 4.0 & 1.2 & -3.2 & y*\\
FM 543$^d$ & 05:46:14.48 & 00:20:24.4 & M4 & 1.8 & 1.4 & -3.6 & n*\\
FM 998$^d$ & 05:46:58.13 & 00:05:38.2 & M0 & 3.7 & 1.2 & -2.0 & y*\\
FM 1056 & 05:47:03.32 & 00:23:23.5 & M3 & 7.5 & 1.7 & \nodata & \nodata \\
OB1b-337$^e$ & 05:29:35.44 & -01:39:38.9 & K7 & 0.0 & 0.93 & -1.3 & n \\
CVSO 103/OB1b-1810$^f$ & 05:32:06.41 & -01:26:43.5 & M2 & 0.0 & 1.5 & -20 & y \\
OB1b-2266$^e$ & 05:32:43.71 & -01:58:10.9 & M3.5 & 0.0 & 1.4 & -3.7 & n \\
OB1a-905$^e$ & 05:24:58.85 & +01:25:18.3 & M5 & 0.0 & 1.3 & -15 & n \\
CVSO 217/OB1a-1121$^f$ & 05:25:34.40 & +01:52:19.7 & M1 & 0.0 & 0.98 & -2.1 & n\\
OB1a-1695$^e$ & 05:27:02.95 & +01:39:00.8 & M5 & 0.0 & 1.0 & -6.7 & n \\
RECX 9$^g$ & 08:44:16.37 & -78:59:08.0 & M4.5 & 0.0 & 1.3 & -10.0 & y*\\
ECHA J0841.5-7853$^g$ & 08:41:30.30 & -78:53:06.4 & M4.75 & 0.0 & 1.4 & -12.0 & n* \\
ECHA J0844.2-7833$^h$ & 08:44:09.14 & -78:33:45.7 & M5.75 & 0.0 & 1.7 & \nodata & \nodata \\
\enddata
\tablecomments{Spectral types and H$\alpha$ measurements from:
$^a$Luhman et al. (2001); $^b$Dahm (2008); $^c$Luhman et al. (2003);
$^d$Flaherty \& Muzerolle (2008); $^e$Brice\~no et al. in preparation;
$^f$Brice\~no et al. (2007); $^g$Lawson et al. (2002);
$^h$Luhman \& Steeghs (2004).
If the extinction was adjusted for a better SED fit (see text),
the original value from the literature is listed in parentheses.
Accretor status is marked as in Table~\ref{known}.}
\end{deluxetable}

\clearpage

\begin{references}

Alexander, R. D., Clarke, C. J., \& Pringle, J. E. 2006, MNRAS, 369, 229

Alexander, R. D. \& Armitage, P. J. 2009, ApJ, 704, 989

Balog, Z., Muzerolle, J., Rieke, G. H., Su, K. Y. L., Young, E. T., \& Megeath, S. T. 2007, ApJ, 660, 1532

Baraffe, I., Chabrier, G., Allard, F., \& Hauschildt, P. H. 1998, \aap, 337, 403

Bergin, E., Calvet, N., Sitko, M. L. et al. 2004, ApJ, 614, L133

Boss, A. P. 1997, Science, 276, 5320, 1836

Bouvier, J. \& Appenzeller, I. 1992, A\&AS, 92, 481

Bouwman, J., Lawson, W. A., Dominik, C., Feigelson, E. D., Henning, T., Tielens, A. G. G. M., Waters, L. B. F. M. 2006, ApJ, 653, L57

Brandeker, A., Jayawardhana, R., Khavari, P., Haisch, K. E., \& Mardones, D. 2006, ApJ, 652, 1572

Brice\~no, C., Calvet, N., Hern\'andez, J., Vivas, A. K., Hartmann, L., Downes, J. J., \& Berlind, P. 2005, AJ, 129, 907

Brice\~no, C., Hartmann, L., Hern\'andez, J., Calvet, N., Vivas, A. K., Furesz, G., \& Szentgyorgyi, A. 2007, ApJ, 661, 1119

Brown, J. M., Blake, G. A., Dullemond, C. P. et al. 2007, ApJ, 664, L107

Brown, J. M., Blake, G. A., Qi, C., Dullemond, C. P., \& Wilner, D. J. 2008, ApJ, 675, L109

Calvet, N., D'Alessio, P., Hartmann, L., Wilner, D., Walsh, A. \& Sitko, M. 2002, ApJ, 568, 1008

Carpenter, J. M., Mamajek, E. E., Hillenbrand, L. A., \& Meyer, M. R. 2006, ApJ, 651, L49

Cieza, L., Padgett, D. L., Stapelfeldt, K. R. \etal 2007, ApJ, 667, 308

Cieza, L. A., Swift, J. J., Mathews, G. S., \& Williams, J. P. 2008, ApJ, 686, L115

Currie, T. \& Kenyon, S. J. 2009a, AJ, 138, 703

Currie, T., Lada, C. J., Plavchan, P., Robitaille, T. P., Irwin, J., \& Kenyon, S. J. 2009b, ApJ, 698, 1

D'Alessio, P., Calvet, N., Hartmann, L., Lizano, S., \& Cant\'o, J. 1999, ApJ, 527, 893

D'Alessio, P., Hartmann, L., Calvet, N. et al. 2005, ApJ, 621, 461

D'Alessio, P., Calvet, N., Hartmann, L., Franco-Hern\'andez, R., \& Serv\'in, H. 2006, ApJ, 638, 314

Dahm, S. E. 2008, AJ, 136, 521

Dahm, S. E. \& Carpenter, J. M. 2009, AJ, in press

Dahm, S. E. \& Hillenbrand, L. A. 2007, AJ, 133, 2072

Duch\^ene, G., Bouvier, J., \& Simon, T. 1999, A\&A, 343, 831

Dullemond, C. P. \& Dominik, C. 2005, A\&A, 434, 971

Eisner, J. A., Chiang, E. I., \& Hillenbrand, L. A. 2006, ApJ, 637, L133

Ercolano, B., Clarke, C. J., \& Robitaille, T. P. 2009, MNRAS, in press

Espaillat, C. et al. 2007a, ApJ, 664, L111

Espaillat, C. et al. 2008a, ApJ, 682, L125

Espaillat, C. et al. 2008b, ApJ, 689, L145

Flaherty, K. M. \& Muzerolle, J. 2008, AJ, 135, 966

Flaherty, K. M., Pipher, J. L., Megeath, S. T., Winston, E. M., Gutermuth, R. A., Muzerolle, J., Allen, L. E., \& Fazio, G. G. 2007, ApJ, 663, 1069

Furlan, E. et al. 2006, ApJS, 165, 568

Furlan, E. et al. 2007, ApJ, 664, 1176

Furlan, E., et al. 2009, ApJ, 703, 1964

Gautier, T. N., Rebull, L. M., Stapelfeldt, K. R., \& Mainzer, A. 2008, ApJ, in press

Gutermuth, R. A., Megeath, S. T., Muzerolle, J., Allen, L. E., Pipher, J. L., Myers, P. C., \& Fazio, G. G. 2004, ApJS, 154, 374

Gutermuth, R., Megeath, S. T., Pipher, J. L., Williams, J. P., Allen, L. E., Myers, P. C., \& Raines, S. N. 2005, ApJ, 632, 397

Gutermuth, R., Myers, P. C., Megeath, S. T. et al. 2008, ApJ, 674, 336

Hartmann, L., Megeath, S. T., Allen, L. E., Luhman, K., Calvet, N., D'Alessio, P., Franco-Hernandez, R., \& Fazio, G. 2005, ApJ, 629, 881

Hern\'andez, J., Calvet, N., Brice\~no, C., et al. 2007, ApJ, 671, 1784

Hern\'andez, J., Hartmann, L., Megeath, S. T., et al. 2007, ApJ, 662, 1067

Hillenbrand, L. 2005, review article in ``A Decade of Discovery: Planets Around Other Stars", STScI Symposium Series 19, ed. M. Livio

Hughes, A. M., Andrews, S. M., Espaillat, C., et al. 2009, ApJ, in press

Hughes, A. M., Wilner, D. J., Calvet, N., D'Alessio, P., Claussen, M. J., \& Hogerheijde, M. R. 2007, ApJ, 664, 536

Ireland, M. J. \& Kraus, A. L. 2008, ApJ, 678, L59

Kennedy, G. M. \& Kenyon, S. J. 2009, ApJ, in press

Kenyon, S. J. \& Hartmann, L. 1995, ApJS, 101, 117

Kroupa, P. 2001, MNRAS, 322, 231

Lada, C. J., Muench, A. A., Luhman, K. L. et al. 2006, AJ, 131, 1574

Lawson, W. A., Crause, L. A., Mamajek, E. E., \& Feigelson, E. D. 2002, MNRAS, 329, L29

Lawson, W. A., Lyo, A.-R., \& Muzerolle, J. 2004, MNRAS, 351, L39

Loinard, L., Torres, R. M., Mioduszewski, A. J., \& Rodr\'iguez, L. F. 2008, ApJ, 675, L29

Low, F. J., Smith, P. S., Werner, M., Chen, C., Krause, V., Jura, M., Hines, D. C. 2005, ApJ, 631, 1170

Luhman, K. L. \& Rieke, G. H. 1999, ApJ, 525, 440

Luhman, K. L., Stauffer, J. R., Muench, A. A., Rieke, G. H., Lada, E. A., Bouvier, J., \& Lada, C. J. 2003, \apj, 593, 1093

Luhman, K. L. \& Steeghs, D. 2004, ApJ, 609, 917

Mathis, J. S. 1990, ARA\&A, 28, 37

McCabe, C., Ghez, A. M., Prato, L., Duch\^ene, G., Fisher, R. S., \& Telesco, C. 2006, ApJ, 636, 932

Megeath, S. T., Hartmann, L., Luhman, K. L., \& Fazio, G. G. 2005, ApJ, 634, L113

Mitchell, G. F., Johnstone, D., Moriarity-Schieven, G., Fich, M., \& Tothill, N. F. H. 2001, ApJ, 556, 215

Moraux, E., Lawson, W. A., \& Clarke, C. 2007, A\&A, 473, 163

Muzerolle, J., Calvet, N., Hartmann, L., D'Alessio, P. 2003, ApJ, 597, L149

Padgett, D. L., Cieza, L., Stapelfeldt, K. R. et al. 2006, ApJ, 645, 1283

Pollack, J. B., Hubickyj, O., Bodenheimer, P., Lissauer, J. J., Podolak, M., \& Greenzweig, Y. 1996, Icarus, 124, 63

Quillen, A. C., Blackman, E. G., Frank, A., \& Varni\`ere, P. 2004, ApJ, 612, L137

Ratzka, T., Leinert, C., Henning, T., Bouwman, J., Dullemond, C. P., \& Jaffe, W. 2007, A\&A, 471, 173

Rice, W. K. M., Armitage, P. J., Wood, K., \& Lodato, G. 2006, MNRAS, 373, 1619

Rice, W. K. M., Wood, K., Armitage, P. J., Whitney, B. A., \& Bjorkman, J. E. 2003, MNRAS, 342, 79

Sicilia-Aguilar, A., Hartmann, L., Brice\~no, C., Muzerolle, J., \& Calvet, N. 2004, AJ, 128, 805

Sicilia-Aguilar, A., Hartmann, L., Calvet, N. et al. 2006, ApJ, 638, 897

Sicilia-Aguilar, A., Henning, T., Juh\'asz, A., Bouwman, J., Garmire, G., \& Garmire, A. 2008, ApJ, 687, 1145

Siess, L., Dufour, E., \& Forestini, M. 2000, A\&A, 358, 593

Skrutskie, M. F., Dutkevitch, D., Strom, S. E., Edwards, S., \& Strom, K. M. 1990, AJ, 99, 1187

Sterzik, M. E., Alcala, J. M., Covino, E., \& Petr, M. G. 1999, A\&A, 346, L41

Strom, K. M., Strom, S. E., Edwards, S., Cabrit, S., \& Skrutskie, M. F. 1989, AJ, 97, 1451

Uchida, K. I., Calvet, N., Hartmann, L. et al. 2004, ApJS, 154, 439

Webb, R. A. et al. 1999, ApJ, 512, L63

Weidenschilling, S. J. 1997, Icarus, 127, 290

Wilking, B. A., Meyer, M. R., Greene, T. P., Mikhail, A., \& Carlson, G. 2004, AJ, 127, 1131

Winston, E., Megeath, S. T., Wolk, S. J. et al. 2007, ApJ, 669, 493

Wood, K., Lada, C. J., Bjorkman, J. E., Kenyon, S. J., Whitney, B., \& Wolff, M. J. 2002, ApJ, 567, 1183

Zuckerman, B., Webb, R. A., Schwartz, M., \& Becklin, E. E. 2001, ApJ, 549, L233

\end{references}
\end{document}